\documentclass[prl,twocolumn,superscriptaddress,notitlepage]{revtex4-1}
\usepackage{makeidx}
\usepackage[utf8]{inputenc}
\usepackage{graphicx}
\usepackage{amsmath}
\usepackage{mathrsfs}
\usepackage{graphicx}
\usepackage{subfigure}
\usepackage{dcolumn}
\usepackage{bm}
\usepackage{bbm}
\usepackage{color}
\usepackage{esint}
\usepackage{lineno}
\usepackage[breaklinks,
            colorlinks,
            urlcolor=blue,
            linkcolor=blue,
            anchorcolor=blue,
            citecolor=blue]{hyperref}
\usepackage[dvipsnames]{xcolor}
\usepackage{overpic}
\usepackage{hyperref}

\renewcommand{\section}[1]{{\par\it #1.---}\ignorespaces}

\begin{document}

\title{Probing dynamics of time-varying media: Beyond abrupt temporal interfaces}

\author{Ayan Nussupbekov}
\email{nussupbekov\_ayan@ihpc.a-star.edu.sg}
\affiliation{
Institute of High Performance Computing, A*STAR (Agency for Science, Technology, and Research), 1 Fusionopolis Way, No. 16-16 Connexis, Singapore 138632.
}

\author{Juan-Feng Zhu}
\email{juanfeng\_zhu@sutd.edu.sg}
\affiliation{Science, Mathematics, and Technology (SMT), Singapore University of Technology and Design (SUTD), 8 Somapah Road, Singapore 487372.}

\author{Yuriy Akimov}
\affiliation{
Institute of High Performance Computing, A*STAR (Agency for Science, Technology, and Research), 1 Fusionopolis Way, No. 16-16 Connexis, Singapore 138632.
}

\author{Ping Bai}
\affiliation{
Institute of High Performance Computing, A*STAR (Agency for Science, Technology, and Research), 1 Fusionopolis Way, No. 16-16 Connexis, Singapore 138632.
}
\author{Ching Eng Png}
\affiliation{
Institute of High Performance Computing, A*STAR (Agency for Science, Technology, and Research), 1 Fusionopolis Way, No. 16-16 Connexis, Singapore 138632.
}

\author{Francisco J. Garcia-Vidal}
\affiliation{%
Departamento de Fisica Teorica de la Materia Condensada and Condensed Matter Physics Center (IFIMAC), Universidad Autonoma de Madrid, E-28049 Madrid, Spain.
} 
\affiliation{
Institute of High Performance Computing, A*STAR (Agency for Science, Technology, and Research), 1 Fusionopolis Way, No. 16-16 Connexis, Singapore 138632.
}

\author{Lin Wu}
\email{lin\_wu@sutd.edu.sg}
\affiliation{Science, Mathematics, and Technology (SMT), Singapore University of Technology and Design (SUTD), 8 Somapah Road, Singapore 487372.}
\affiliation{
Institute of High Performance Computing, A*STAR (Agency for Science, Technology, and Research), 1 Fusionopolis Way, No. 16-16 Connexis, Singapore 138632.
}

\begin{abstract}
This work investigates the effects of time-varying media, where optical properties change over time, on electromagnetic wave propagation, focusing on plane waves and free-electron evanescent waves. We introduce a switching parameter, $\tau$, to model ultrafast switching in the femtosecond to nanosecond range. For plane-wave incidence at angular frequency $\omega_0$, we derive a generalized expression for the backward-to-forward flux ratio as a function of $\omega_0$ and $\tau$, aligning with recent experimental data and providing a unified interpretation framework. For free-electron incidence, we observe intensity saturation in temporal transition radiation at 
$I_{\textrm{max}}$ for $\tau \leq \tau_{\textrm{0}}$, with both $I_{\textrm{max}}$ and $\tau_{\textrm{0}}$ depending on electron speed. These results highlight the importance of precise 
$\tau$ control in experiments to probe time-varying media effectively.
\end{abstract}

\maketitle
The principles of reflection and refraction of light have been well understood for centuries, with foundational experiments in optics shaping modern understanding \cite{pedrotti2018introduction}. Over time, these principles have evolved into complex engineering challenges, particularly in the design of hyperbolic media \cite{li2016hyperbolic, poddubny2013hyperbolic}, as well as anisotropic and nonlinear media \cite{saleh2019fundamentals, boyd2008nonlinear}. Recently, there has been increasing interest in temporal phenomena, where the dielectric properties of a material change over time, while remaining spatially uniform \cite{galiffi2022photonics, caloz2019spacetime,mendoncca2002time,zhou2020broadband,lerosey2004time}.

Theoretical studies of temporally varying media date back to the mid-20$^{\textrm{th}}$ century \cite{morgenthaler1958velocity, fante1971transmission, fante1973propagation}, but recent advances have sparked significant experimental interest. Notable developments include temporal reflection, where light interacts with a boundary defined by a sudden change in material properties over time \cite{jones2024time, moussa2023observation,wang2023metasurface}, temporal interference, which extends the classic double-slit experiment into the time domain \cite{tirole2023double}, and second harmonic generation in time-varying systems \cite{tirole2024second}. These phenomena have opened exciting new avenues for research in time-dependent optics \cite{bar2024long,konforty2024second,lyubarov2024controlling,bar2024time,pan2023superluminal,wang2024expanding,sharabi2022spatiotemporal,dikopoltsev2022light}.

While the results from these experiments are intriguing, one key factor that is often overlooked is the \textbf{finite switching time} of the material, which varies significantly across different platforms. For example, nonlinear materials can exhibit permittivity switching on the femtosecond (fs) scale due to the Kerr effect \cite{tirole2023double,guo2019nonreciprocal}, whereas capacitance modulation in transmission lines typically operates on the nanosecond (ns) scale \cite{reyes2015observation,reyes2016electromagnetic}, with many other experimental systems falling between these time scales \cite{wang2023metasurface,sisler2024electrically,zhang2018space,li2014ultrafast,wilson2018temporal,phare2015graphene,wang2023manipulations,ding2024electrically,jung2024rise}. These variations highlight the importance of accounting for switching times when designing experiments and interpreting results.

Most theoretical models \cite{morgenthaler1958velocity,galiffi2022photonics} and recent experiments \cite{moussa2023observation,jones2024time,tirole2023double,tirole2024second,wang2023metasurface} assume an \textbf{instantaneous} change in material properties, neglecting the switching time relative to the electromagnetic wave's period. While this assumption is often reasonable, few theoretical studies have investigated how the temporal dynamics of the switching process, combined with the wavelength of the incident wave, influence the observed effects, including the emergence of instantaneous frequency shifts in both dispersive and non-dispersive media \cite{fante1971transmission}, energy conservation in continuously varying media \cite{hayrapetyan2016electromagnetic}, conditions for total reflection in materials with smooth temporal gradients \cite{zhang2021impact}, and the design of temporal photonic tapers \cite{galiffi2022tapered}.

In this Letter, we highlight the critical role of switching time in material transitions, extending beyond plane-wave incidence to include free electron interactions. We introduce the concept of temporal transition radiation ($t$-TR), which enhances our understanding of how temporal effects depend on switching time, wavelength, and electron speed. We derive a generalized expression for the backward-to-forward flux ratio as a function of switching time, showing strong agreement with recent experimental data and offering a unified interpretation framework. Full-wave simulations validate our model, providing insights into reflection, transmission, and electric field distributions in time-varying media. Furthermore, we demonstrate how switching time influences radiation at the temporal boundary, with results indicating that the requirement for shorter switching times relaxes with higher electron speeds. These findings suggest new opportunities for exploring temporal phenomena over extended timescales. Ultimately, precisely engineered temporal boundaries could enable tailored $t$-TR effects, such as coherent $t$-TR, paving the way for new experimental approaches to probing and controlling switching dynamics.

\begin{figure*}
    \centering
    \begin{overpic}[width = 0.8\textwidth]{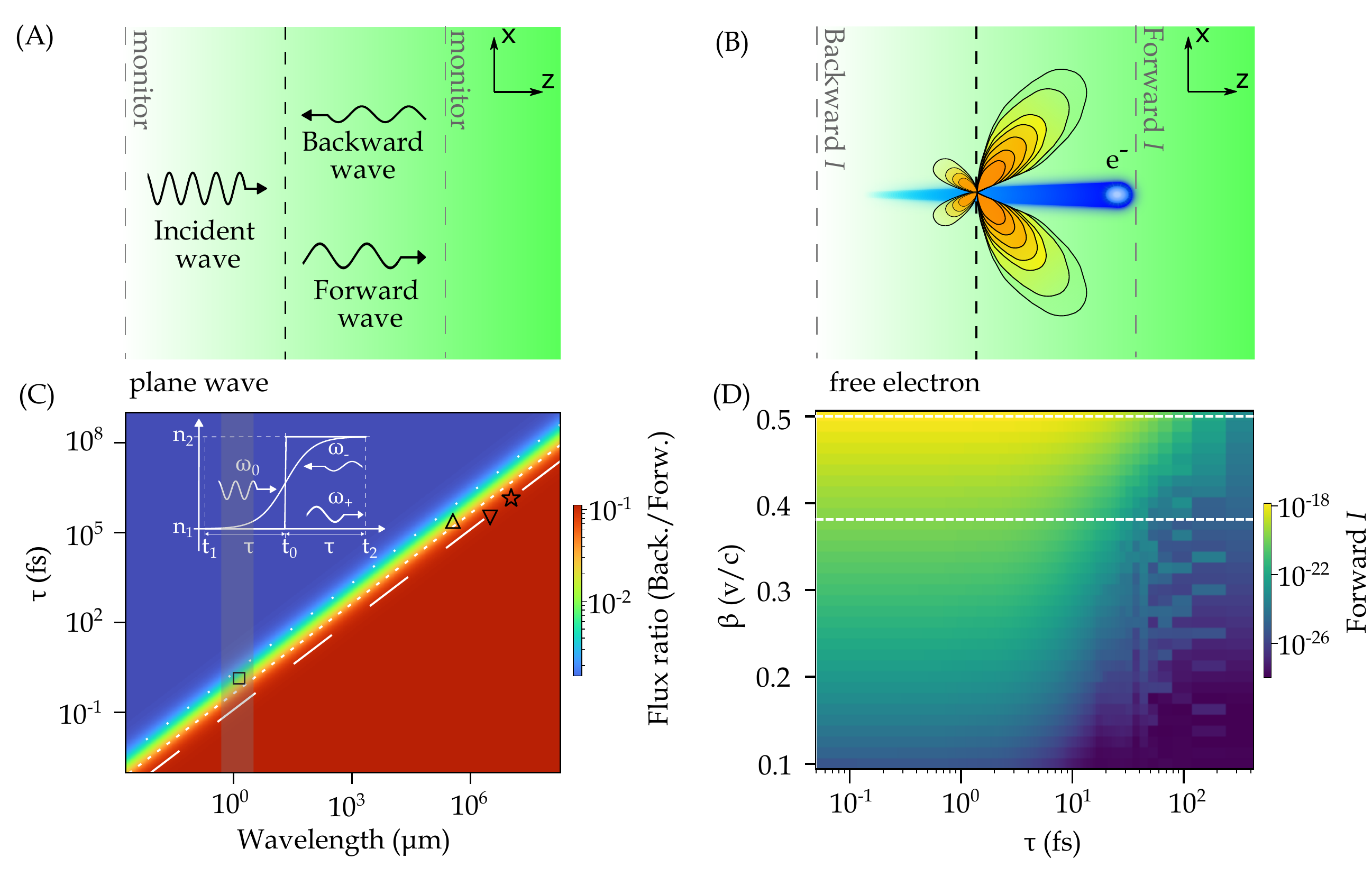}
        \put(18, 13.5){\tiny \textcolor{blue}{\cite{tirole2023double}}}
        \put(18, 13.5){\tiny \textcolor{white}{[17]}}
        % Add citation near a data point
        \put(32, 24){\tiny \textcolor{blue}{\cite{jones2024time}}} % Another reference
        \put(32, 24){\tiny \textcolor{white}{[14]}}
        \put(34.5, 24.5){\tiny \textcolor{blue}{\cite{moussa2023observation}}}
        \put(34.5, 24.5){\tiny \textcolor{white}{[15]}}
        % Another reference
        \put(36.8, 25.7){\tiny \textcolor{blue}{\cite{wang2023metasurface}}} 
        \put(36.8, 25.7){\tiny \textcolor{white}{[16]}}
        % % Another reference
    \end{overpic}
    \caption{Interaction of electromagnetic waves with a continuously varying temporal boundary.  
    (A) A plane wave incident on a time-varying medium generates forward (transmitted) and backward (reflected) waves.  
    (B) A free electron generates temporal transition radiation ($t$-TR). 
    (C) Analytical derivation of the backward-to-forward flux ratio as a function of $\tau$ and wavelength for a plane wave with angular frequency $\omega_0$ incident on a time-varying medium. The boundary transition is controlled by the switching parameter $\tau$ (Inset), which governs the refractive index change from $n_1$ to $n_2$. Data points (squares, triangles, stars) correspond to experimental results from \cite{jones2024time, moussa2023observation, wang2023metasurface, tirole2023double}.  The dotted, short-dashed, and long-dashed lines represent $\omega_0 \tau = 4.32$, $\omega_0 \tau = 0.9$, and $\omega_0 \tau = 0.26$, respectively, with the region between the dotted and long-dashed lines indicating the $5\%$ - $90\%$ maximum flux ratio range. (D) Numerical simulation of a free electron interacting with the temporal boundary. The radiated power at the forward monitor for $t$-TR depends on switching time $\tau$  and electron speed $\beta$ within the wavelength range shown in shaded gray in (C). The unit of forward $I$ is the number of photons per electron.
}
\label{fig:fig1}
\end{figure*}
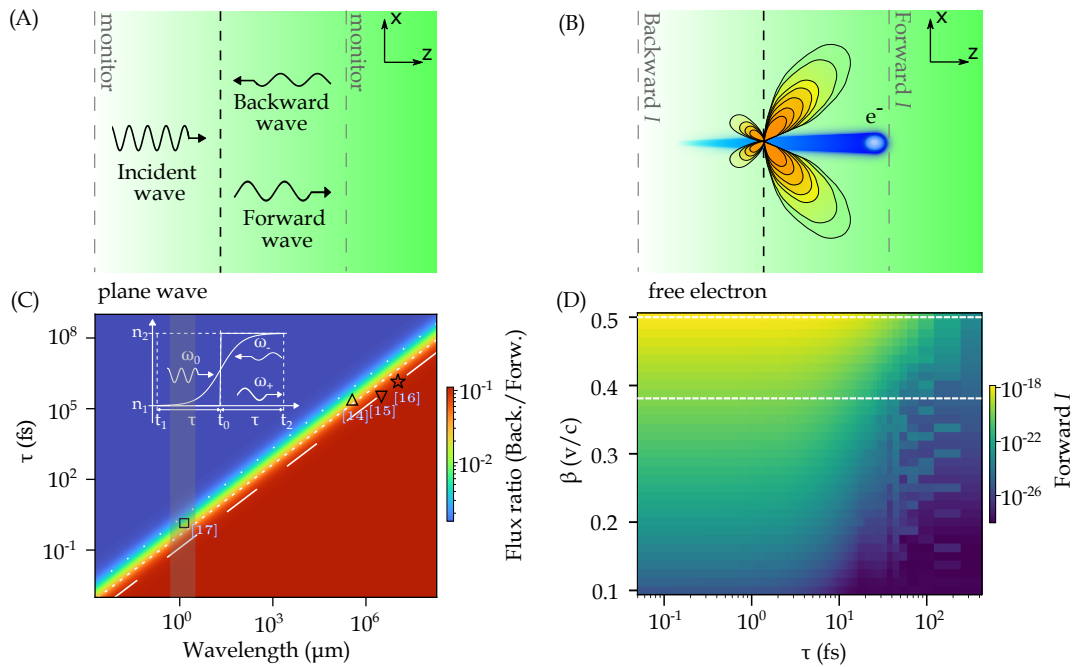

\textcolor{blue}{Figure~\ref{fig:fig1}} offers an overview of electromagnetic (EM) wave interacting with a continuously varying temporal boundary, showcasing two key scenarios: the plane wave (\textcolor{blue}{Fig.~\ref{fig:fig1}A}) and the evanescent waves generated by a free electron (\textcolor{blue}{Fig.~\ref{fig:fig1}B}).

In \textcolor{blue}{Fig.~\ref{fig:fig1}A}, a plane wave propagates through a spatially homogeneous, non-dispersive medium with a time-dependent permittivity $\varepsilon(t)$. When the wave encounters a temporal boundary, it generates both forward (transmitted) and backward (reflected) waves. Unlike a spatial boundary, which features a sharp interface, the temporal boundary lacks a distinct separation and is defined by the transition center, $t_0$ (dashed line), relative to the medium's initial and final refractive indices, $n_1$ and $n_2$ (Inset, \textcolor{blue}{Fig.~\ref{fig:fig1}C}). The switching time, $\tau = |t_{1,2}-t_0|$, quantifies the rate of change in the material's properties and controls how quickly the medium reacts to external influences. This study distinguishes itself from previous works \cite{morgenthaler1958velocity} by considering $\tau$ as a switching parameter, enabling a continuous transition from the abrupt changes typically used in modeling temporal boundaries.

In \textcolor{blue}{Fig.~\ref{fig:fig1}B}, we explore the interaction of a free electron's evanescent fields with a continuously varying temporal boundary. Traditionally, when a charged particle like an electron crosses a spatial boundary, it generates spatial transition radiation ($s$-TR) \cite{transition_radiation}, emitted in both forward and backward directions. $s$-TR has garnered significant attention for its practical applications, as it does not require strict conditions like minimum particle speed or phase matching, making it valuable in particle diagnostics, such as high-energy physics \cite{chen2023recent}.
In analogy, temporal transition radiation ($t$-TR) arises when an electron crosses a temporal boundary. Unlike $s$-TR at a spatial interface, $t$-TR results from the conversion of the electron's evanescent fields into radiating fields at the temporal boundary. Notably, $t$-TR is highly directional, with strong forward emission and minimal backward radiation, a behavior consistent with Ginzburg's theoretical predictions \cite{ginzburg1979several} for radiation at a temporally varying boundary with instantaneous switching [see \textcolor{blue}{SM-I} \cite{supp} for comparison of $s$-TR and $t$-TR].

In \textcolor{blue}{Fig.~\ref{fig:fig1}C} and \textcolor{blue}{Fig.~\ref{fig:fig1}D}, we show how the switching parameter $\tau$ affects EM propagation in two cases: \textcolor{blue}{Fig.~\ref{fig:fig1}C} plots the forward-to-backward flux ratio for a plane wave, while \textcolor{blue}{Fig.~\ref{fig:fig1}D} illustrates the radiated power (forward $I$) for a free electron, measured in photons per electron.

For the plane wave, the flux ratio is plotted as a function of angular frequency $\omega_0$  and switching time $\tau$ in a wide range of parameters. This ratio was analytically derived using Maxwell's equations, with the WKB approximation (details in \textcolor{blue}{SM-II.b} \cite{supp}), yielding the expression:
\begin{align} 
    \frac{|B_-|^2}{|B_+|^2} =
    \frac{\left( n_1 - n_2 \right)^2 + 4 \omega_0^2 \tau^2 \left[ \frac{2 n_1 n_2}{n_2^2 - n_1^2} \ln{\frac{n_2}{n_1}} - 1 \right]^2}{\left( n_1 + n_2 \right)^2 + 4 \omega_0^2 \tau^2 \left[ \frac{2 n_1 n_2}{n_2^2 - n_1^2} \ln{\frac{n_2} {n_1}} + 1 \right]^2}. \label{eq:flux_tau_non_zero} 
\end{align}
Here, $B_+$ and $B_-$ represent the forward and backward-propagating components in the new medium. We set $n_1 = 1$ and $n_2 = 2$, with $\omega_0$ and $\tau$ reflecting recent experimental data \cite{moussa2023observation, jones2024time, tirole2023double, wang2023metasurface}. While $n_1$ and $n_2$ are arbitrary, the trends hold unless the indices differ drastically (see \textcolor{blue}{SM-III} \cite{supp}).

As shown in \textcolor{blue}{Fig.~\ref{fig:fig1}C}, a threshold value of 
$\omega_0 \tau$ exists beyond which the flux ratio saturates, peaking at $1/9$ (dark red region). Minimal effects occur in the dark blue regions, with threshold lines (dotted, short-dashed, and long-dashed) marking the 5\%, 50\%, and 90\% of the peak ratio. For instance, achieving at least 5\% of the peak ratio ($|B_-|^2/|B_+|^2 = 0.005$, white dotted line) requires $\omega_0 \tau \approx 4.32$, meaning the wave period must be approximately 1.5 times the switching time.
Experimental conditions generally fall between the white dotted and long-dashed lines. In the infrared regime (square marker), switching times correspond to 5\%-50\% thresholds \cite{tirole2023double}, reflecting material limitations on rapid transitions. At longer wavelengths \cite{jones2024time, moussa2023observation, wang2023metasurface}, nanosecond switching times align with the 50\%-90\% thresholds. For nearly instantaneous transitions, the wave period must exceed the switching time by a factor of 24 ($|B_-|^2/|B_+|^2 = 0.1$, $\omega_0 \tau \approx 0.26$), favoring longer wavelengths for maximal temporal effects.

\textcolor{blue}{Fig. \ref{fig:fig1}D} shows the calculated forward intensity of $t$-TR as a function of electron speed $\beta$ (normalized to the speed of light) and switching time $\tau$. We perform full-wave FDTD simulations using MEEP \cite{oskooi2010meep} (details in \textcolor{blue}{SM-IV} \cite{supp}), modeling the electron as a series of dipoles \cite{massuda2018smith,lu2023smith} to accurately capture the radiation pattern's dependence on $\beta$ and $\tau$. The forward intensity $I$ is calculated from the Poynting vector, integrated over the monitor line, and averaged over the 0.5 to 3 $\mu\mathrm{m}$ wavelength range (shaded gray region in \textcolor{blue}{Fig.~\ref{fig:fig1}C}), selected to align with ongoing efforts to replicate far-infrared experiments in the near-infrared and visible regimes \cite{tirole2023double,tirole2024second}. These experiments have yet to be explored due to the limitations of time-varying materials \cite{pacheco2022time}.
We observe that at higher electron speeds, the intensity saturates at longer $\tau$, highlighting the increasing influence of temporal effects compared to the plane wave case, with a detailed comparison provided in later sections. This suggests that free electrons are ideal for probing time-varying materials. For this study, $\beta$ was limited to 0.5 to avoid significant Cherenkov radiation (CR) \cite{dikopoltsev2022light, jackson1999classical}, which occurs when the electron’s velocity exceeds the phase velocity of light in the medium. A broader velocity range is explored in \textcolor{blue}{SM-V} \cite{supp}.

\section{Plane wave incident on continuously varying temporal boundary}
While Eq.~(\ref{eq:flux_tau_non_zero}) is most accurate for fast transitions (red regions), it offers valuable insight into flux ratio trends across the $\tau$ spectrum. The slow-transition regime (blue regions) requires more precise modeling.
We begin our analysis by considering a plane wave of angular frequency $\omega_1 = \omega_0 n_1^{-1}$, incident on a non-dispersive medium with an abrupt change, as illustrated in the inset of \textcolor{blue}{Fig.~\ref{fig:fig1}C}. This scenario represents the limiting case of a continuously varying medium when $\tau \rightarrow 0$. 
For this case, Maxwell's equations can be solved for the time intervals before and after the transition at $t_0$. 

Before the transition ($t_1<t_0$), the wave propagates unperturbed, and we assume the harmonic solution for the magnetic field  $\textbf{H}_1 = \mathbf{H}_0 e^{i(\mathbf{k}_1 \cdot \mathbf{r} - \omega_1 t)}$, 
where $\mathbf{H}_0$ is the amplitude and
$\mathbf{k}_1$  is the wave vector. After the transition ($t_2>t_0$), the medium properties change, and the fields become: $\mathbf{H}_2 =\mathbf{H}_0 \left( B_+e^{-i\omega_+ t} + B_-e^{-i\omega_- t} \right) e^{i\mathbf{k}_1 \cdot \mathbf{r}}$.
Here, $\omega_1 $ and $\omega_{\pm} = \pm c k_1 n_2^{-1}$ are the angular frequencies before and after the transition, respectively, where $c$ is the speed of light. Conservation of momentum implies a relationship between the frequencies before and after the transition:
$\omega_1 n_1 = \pm\omega_{\pm}n_2$.

To determine the coefficients $B_+$ and $B_-$, we match the boundary conditions of the incident ($\mathbf{E}_1, \mathbf{H}_1$) and scattered ($\mathbf{E}_2, \mathbf{H}_2$) fields at $t_0 = 0$.  This yields the ratio of backward to forward wave fluxes (details in \textcolor{blue}{SM-II.a} \cite{supp}):
\begin{equation}
    \frac{|B_-|^2}{|B_+|^2} = \left( \frac{n_1 - n_2}{n_1 + n_2} \right)^2.
    \label{eq:flux_tau_zero}
\end{equation}
This expression aligns with similar results in previous works  \cite{hayrapetyan2016electromagnetic,galiffi2022photonics,morgenthaler1958velocity}. Notably, Eq. (\ref{eq:flux_tau_non_zero}) simplifies to this form for instantaneous transitions ($\omega_0\tau = 0$).

Equations (\ref{eq:flux_tau_non_zero}) and (\ref{eq:flux_tau_zero}) are approximations of a more general case. While Eq. (\ref{eq:flux_tau_non_zero}) applies to fast transitions, it becomes less accurate for slow transition regimes. The more general expression, detailed in \textcolor{blue}{SM-II.c} \cite{supp}, is:
\begin{align} 
    \frac{|B_-|^2}{|B_+|^2} =
    \frac{\left| D_{-}^{1} D^{4} - D^{2 }D^{3}_{-} \right|^2}{\left| D_{+}^{1}D^{4} - D^{2}D_{+}^{3} \right|^2}. 
    \label{eq:flux_tau_general} 
\end{align}
Here $D$ terms are defined as:
\begin{align*} 
    D_{\pm}^{1} &= Y_0(\xi_2) \pm is Y'_0(\xi_2) \\
    D^{2} &=  Y_0(\xi_1) - is Y'_0(\xi_1) \\
    D_{\pm}^{3} &= J_0(\xi_2) \pm is J'_0(\xi_2) \\
    D^{4} &=  J_0(\xi_1) - is J'_0(\xi_1), 
    \label{eq:D} 
\end{align*}
where $J_0'$ and $Y_0'$ are Bessel functions of the first and second kind, their time derivatives, and $s = sign(n_2^2 - n_1^2)$, $\xi_{1,2} = 4n_{1,2} |\omega_0 \tau/ (n_2^2 - n_1^2)|$.

\begin{figure}
\centering
\includegraphics[width=0.47\textwidth]{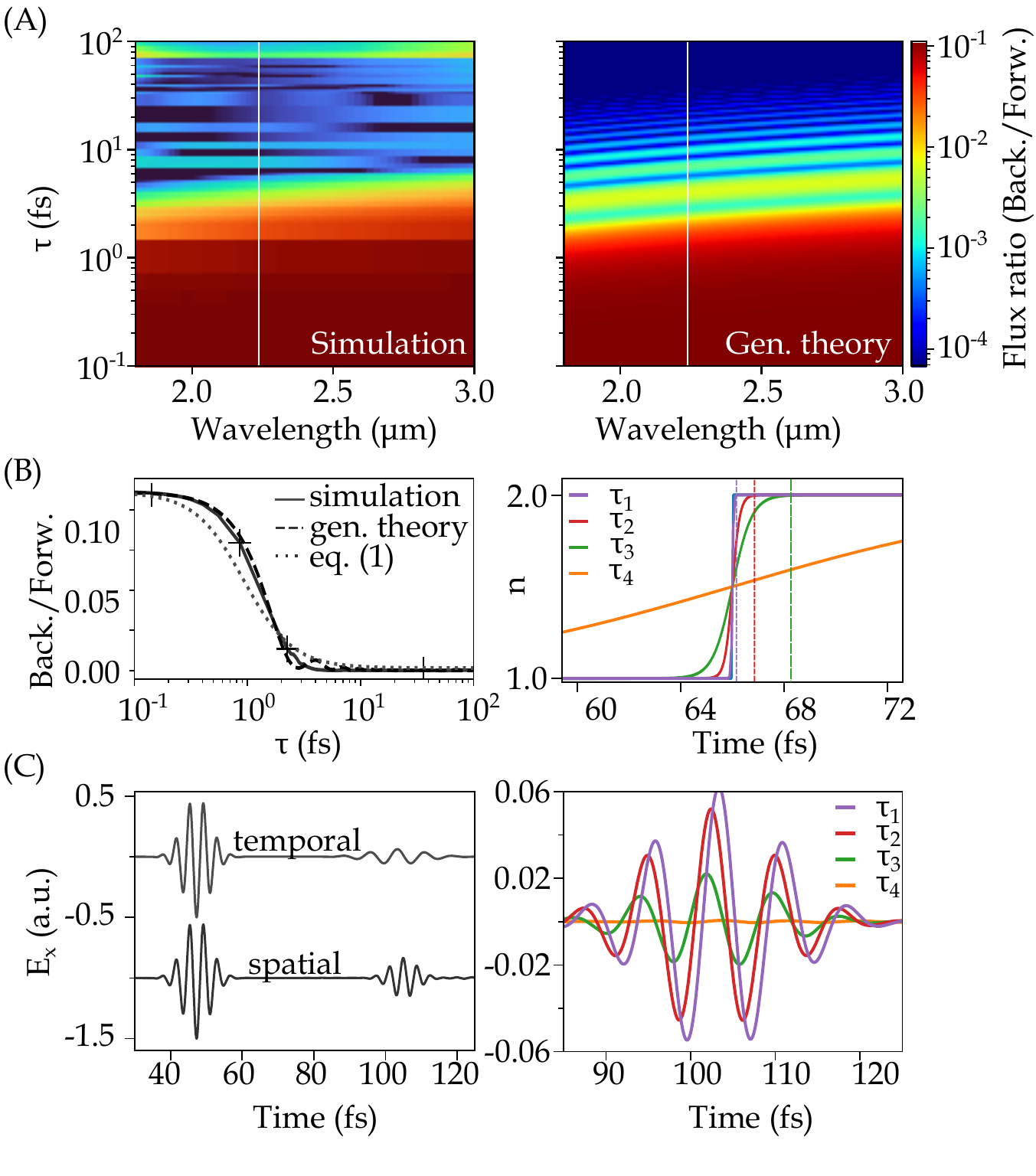}
\caption{
Analysis of a plane wave incident on a time-varying boundary.
(A) Left: FDTD simulation results; right: Theoretical predictions based on Maxwell's equations.
(B) Left: Comparison at a wavelength of 2.3 $\mu$m (white line in (A)); right: Refractive index change profiles at four different switching parameter values (marked by crosses). 
(C) Left: $y$-component of the electric field at the reflectance monitor, showing temporal and spatial reflection; right: Magnified views highlight signal strength variations corresponding to the switching parameter values in (B).
}
\label{fig:fig2}
\end{figure}

To validate the general theory in Eq.~(\ref{eq:flux_tau_general}), we perform MEEP FDTD simulations, as detailed in \textcolor{blue}{SM-IV} \cite{supp}. The left panel of \textcolor{blue}{Fig.~\ref{fig:fig2}A} shows the simulated flux ratio for various switching times $\tau$ and incident wavelengths. The right panel presents the corresponding theoretical results from Eq.~(\ref{eq:flux_tau_general}), demonstrating strong agreement between simulation and general theory. 
A closer inspection of a constant-wavelength slice in \textcolor{blue}{Fig.~\ref{fig:fig2}B} further confirms the strong agreement between simulations and theory across various switching times $\tau$. While the special case (Eq.~(\ref{eq:flux_tau_non_zero})) shown in \textcolor{blue}{Fig.~\ref{fig:fig1}C} deviates from the simulations, the general theory captures the overall trend, especially for fast ($\tau_1$) and intermediate ($\tau_2$ and $\tau_3$) switching rates. The slight discrepancies in the slow-switching regime are likely due to the limited sensitivity of the simulations, which may fail to detect weaker signals.

Full-wave simulations also enable comparison of the electric field for temporal and spatial reflectance (details in \textcolor{blue}{SM-V} \cite{supp}), shown in the left panel of \textcolor{blue}{Fig.~\ref{fig:fig2}C}, where $E_x$ is plotted. The same parameters are used for both cases, with the temporal case corresponding to a fast transition ($\tau_1$). The first peak, representing the incident wavelength, is identical for both cases, while the second peak around 100 fs is broader and weaker in the temporal case. This trend, observed in previous studies \cite{galiffi2022photonics,wang2023metasurface}, results from an increase in the effective refractive index, which redshifts the central frequency via $\omega_{\pm} = \pm \frac{\sqrt{\varepsilon_1}}{\sqrt{\varepsilon_2}}\omega_1$. Additionally, the field amplitude decreases due to the Coulomb effect, described by $\mathbf{E}(t > t_0) \sim \mathbf{E}(t < t_0) \cdot \varepsilon_1 / \varepsilon_2$.
The right panel of \textcolor{blue}{Fig.~\ref{fig:fig2}C} shows that the amplitude of the secondary peak decreases as the transition time $\tau$ increases, consistent with results in \textcolor{blue}{Fig.~\ref{fig:fig2}B}. This amplitude reduction scales with $\omega_0 \tau$, highlighting the importance of optimizing the $\omega_0 \tau$ product to enhance temporal effects. For \textcolor{blue}{Fig.~\ref{fig:fig2}C}, at least half the performance of an instantaneous change requires $\omega_0 \tau \approx 2.5$.

\begin{figure}
\centering
\includegraphics[width = 0.43\textwidth]{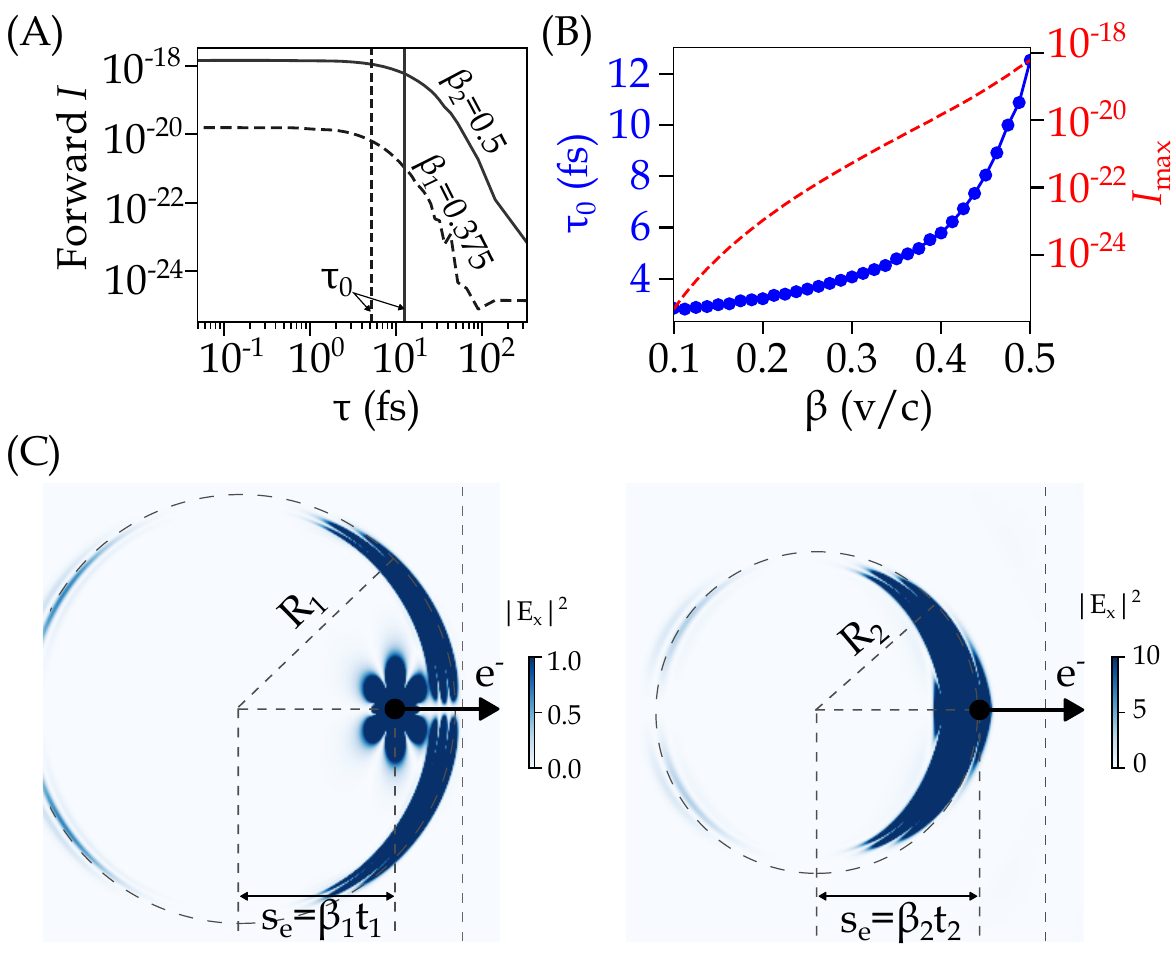}
\caption{Analysis of $t$-TR. (A) Power collected at the forward monitor for $t$-TR as a function of switching parameter $\tau$ at two electron speeds $\beta$ (white lines in Fig. 1B), where $\tau_0$ indicates the threshold for observing maximum or saturated $t$-TR power. (B) Plot showing $\tau_0$ and maximum or saturated $t$-TR power $I_{\textrm{max}}$ as a function of $\beta$. (C) The radiation patterns of $t$-TR at two $\beta$ values show radii $R_1$ and $R_2$, representing the distances traveled by the $t$-TR wave. Two field distributions are plotted at the same electron travel distance $s_e = \beta_1 t_1 = \beta_2 t_2$ from the temporal boundary (central dashed line). The radius difference ($R_1 > R_2$) arises from $t$-TR traveling faster than the electron ($\beta_1 < c/n_2$) in the left panel, and both traveling at the same speed in the right panel ($\beta_2 = c/n_2$).}
\label{fig:fig3}
\end{figure}

\section{Free electron incident on a continuously varying temporal boundary}
Building on the theoretical foundation, we now examine evanescent waves generated by a moving electron. As discussed in \textcolor{blue}{Fig.~\ref{fig:fig1}}, the interaction of a free electron with a temporal boundary produces 
$t$-TR. For a continuously varying boundary, we observe intensity saturation at higher 
$\tau$ values.
This trend is further illustrated in \textcolor{blue}{Fig.~\ref{fig:fig3}A}, showing the intensity of $t$-TR for two electron speeds, marked by white dashed lines in \textcolor{blue}{Fig.~\ref{fig:fig1}D}. As 
$\tau$ decreases, the forward intensity 
$I$ rises sharply, leveling off at a plateau value $I_{\textrm{max}}$ at a characteristic switching time $\tau_0$. 
Taking $\beta = 0.375$ as an example, the forward $t$-TR intensity steadily increases as the switching speed of the medium improves. However, it eventually plateaus at $I_{\textrm{max}} = 10^{-20}$, with this saturation occurring consistently when the switching time remains below $\tau_0 = 5~\mathrm{fs}$. 
A comparison with the plane wave case is revealing. As shown in \textcolor{blue}{Fig.~\ref{fig:fig1}B}, the flux ratio for a plane wave of 0.5--3 $\mu\mathrm{m}$ saturates around $10^{-1}\mathrm{fs}$, an order of magnitude lower than in the free electron case. 
Examining $\tau_0$ and $I_{\textrm{max}}$ as functions of the normalized electron speed $\beta$ in \textcolor{blue}{Fig.~\ref{fig:fig3}B}, we find that saturation occurs for $\beta < 0.1$. This highlights a key insight: free electrons are more effective at probing time-varying media at speeds an order of magnitude lower than those of a plane wave. This advantage may stem from the evanescent, broadband nature of the free electron.

The radiation patterns of $t$-TR for two different electron velocities $\beta$ are shown in \textcolor{blue}{Fig.~\ref{fig:fig3}C}. The $t$-TR radiation is symmetrically emitted around the electron’s trajectory as it crosses the temporal boundary. 
In both panels, $s_e$ denotes the distance traveled by the electron after interacting with the temporal boundary, given by $s_e = \beta t_e$, where $t_e$ is the elapsed time since interaction. The radii $R_1$ and $R_2$ correspond to the wavefront positions of the $t$-TR radiation for electron velocities $\beta_1$ and $\beta_2$, respectively.
The left panel shows the case where $\beta_1 < c/n_2$, where the phase velocity of the emitted wavefront exceeds the electron’s velocity, resulting in a larger radius $R_1$. The right panel shows $\beta_2 = c/n_2$, where the electron velocity matches the phase velocity, producing a smaller radius $R_2$. The intensity pattern is similar, but the angular distribution broadens slightly, indicating stronger interaction with the temporal boundary due to the higher $\beta$.
This difference in radii and angular distribution highlights how $t$-TR radiation patterns depend on electron speed $\beta$ and refractive index $n_2$, similar to $s$-TR. Faster electrons produce broader, more intense radiation patterns, while slower electrons generate more compact wavefronts.

When the electron speed exceeds the Cherenkov radiation (CR) threshold ($\beta \geq 0.5$) \cite{jackson1999classical}, $\tau_0$ increases exponentially, marking the transition from a pure $t$-TR regime to a mixed $t$-TR and CR regime (details in \textcolor{blue}{SM-VI} \cite{supp}). The field patterns for speeds above the CR threshold are shown in the right panel of \textcolor{blue}{Fig.~\ref{fig:fig3}C}. Unlike the pure $t$-TR case in the left panel, the CR regime exhibits stronger electric fields and initial cone formation. However, the small CR angle at the threshold speed ($\cos\theta = 1/(n\beta)$) makes it hard to distinguish from $t$-TR.

\section{Conclusion}
In this study, we investigated electromagnetic wave dynamics interacting with temporal boundaries in time-varying media, introducing a switching parameter, $\tau$, to model continuous transitions. We found that for significant temporal effects in the plane wave case, the wave period must be at least 1.5 times longer than $\tau$, and for instantaneous transitions, this threshold increases dramatically to 24 times the switching time. While nanosecond-scale switching in the longer-wavelength regime can meet these conditions, sub-femtosecond or attosecond switching is required for similar effects in the near-infrared or visible ranges, highlighting the experimental challenges of achieving fast transitions with plane waves. In contrast, $t$-TR provides a more practical method for probing time-varying materials. Our simulations show that $t$-TR exhibits strong forward directionality, with intensity increasing with electron speed and shorter switching times. Importantly, $t$-TR in the near-infrared does not require extremely fast transitions, even at low electron velocities, making it promising for future applications. These results provide a framework for experiments and applications involving temporal boundaries, with potential in particle diagnostics and radiation sources, and offer new possibilities for high-speed applications in time-modulated materials.

\subsection{Acknowledgements}
\begin{acknowledgments}
This work was supported by the National Research Foundation Singapore (Grants NRF2021-QEP2-02-P03, NRF2021-QEP2-03-P09, NRF-CRP26-2021-0004), the Ministry of Education Singapore (Grant MOE-T2EP50223-0001), and the Singapore University of Technology and Design (Kickstarter Initiatives SKI 2021-02-14, SKI 2021-04-12).
\end{acknowledgments}

\bibliography{biblio}% Produces the bibliography via BibTeX.

\end{document}